\documentclass[twocolumn,showpacs,preprintnumbers,amsmath,amssymb]{revtex4}

\usepackage{CJK}
\usepackage{graphicx}
\usepackage{dcolumn}
\usepackage{bm}

\def\btbl{\begin{tabular}} \def\etbl{\end{tabular}}
\def\bcc{\begin{center}} \def\ecc{\end{center}}
\def\beq{\begin{equation}} \def\eeq{\end{equation}}
\def\btbl{\begin{tabular}} \def\etbl{\end{tabular}}

\def\E941{{\footnotesize E941}} \def\E864{{\footnotesize E864}}
\def\NA49{{\footnotesize NA49}} \def\NA35{{\footnotesize NA35}}

\begin{document}
\title{The spatial distributions of magnetic field  in the RHIC and LHC energy regions }

\author{Yang Zhong$^{1,2}$}
\author{Chun-Bin Yang$^{1,3}$}
\author{Xu Cai$^{1,3}$}
\author{Sheng-Qin Feng$^{2,3}$}

\affiliation{$^1$ Institute of Particle Physics, Central China Normal University, Wuhan 430079, China}
\affiliation{$^2$ Department of Physics, College of Science, China Three Gorges University, Yichang 443002, China}
\affiliation{$^3$ Key Laboratory of Quark and Lepton Physics (MOE), Central China Normal University, Wuhan 430079, China}

\begin{abstract}
Relativistic heavy-ion collisions can produce extremely strong  magnetic field in the collision regions.
The spatial variation features of the magnetic fields are analyzed in detail for non-central
Pb - Pb  collisions at  LHC $\sqrt{s_{NN}}$= 900, 2760 and 7000 GeV and
Au-Au collisions at RHIC  $\sqrt{s_{NN}}$ = 62.4, 130 and 200 GeV. The dependencies of magnetic field  on proper time, collision energies
and impact parameters are investigated in this paper. It is shown that a enormous with highly inhomogeneous spatial distribution magnetic field can
indeed be created in off-central relativistic heavy-ion collisions in RHIC and LHC energy regions. The enormous magnetic field is produced just after the collision, and the magnitude of magnetic field of LHC energy region is larger that of RHIC energy region at the small proper time. It is found that the magnetic field in the LHC energy region decreases more quickly with the increase of the proper time than that of RHIC energy region. \\
\vskip0.2cm \noindent Keywowds: Spatial distribution of chiral magnetic field, ~Non-central collision, chiral magnetic field
\end{abstract}

 \maketitle

\section{Introduction}
\label{intro}

The Chiral Magnetic Effect (CME) is the phenomenon of electric charge separation along the
external magnetic field that is introduced by the chirality imbalance ~\cite{lab1,lab2}.
It is proposed by Ref.~\cite{lab3,lab4,lab5,lab6,lab7} that off-central
relativistic heavy-ion collisions can create strong transient
magnetic fields due to the fast, oppositely directed motion of two colliding
nuclei. The magnetic field  perpendicular to the reaction plane is aligned.
Extremely strong (electromagnetic) magnetic fields are present in non-central
collisions, albeit for a very short time.  Thus, relativistic heavy-ion collisions provide a unique terrestrial
environment to study QCD in strong magnetic field  surroundings~\cite{lab8,lab9,lab10,lab11}.
This so-called chiral magnetic effect may serve as a sign of the
local P and CP violation of QCD. By using relativistic heavy-ion collisions at the Relativistic Heavy-Ion Collider (RHIC) and the Large
Hadron Collider (LHC), one can investigate the behavior of QCD at extremely high-energy densities.

In non-central collisions opposite charge quarks would tend to be emitted in opposite directions
relative to the system angular momentum~\cite{lab9,lab12,lab13,lab14}. This asymmetry in the emission of quarks would be reflected in
an analogous asymmetry between positive- and negative-pion emission directions. This phenomenon is introduced by the large
(electro-) magnetic field produced in non-central heavy-ion collisions.  The same phenomenon can also be depicted in terms
of induction of electric field by the (quasi) static magnetic field,
which happens in the occurrence of these topologically nontrivial vacuum solutions. The induced electric field is parallel
to the magnetic field and leads to the charge separation in that direction. Thus, the charge separation can be viewed as a
nonzero electric dipole moment of the system.

Experimentally, RHIC ~\cite{lab15,lab16,lab17,lab18,lab19} and LHC ~\cite{lab20} have
published the measurements of CME by the two-particle or three-particle correlations of
charged particles with respect to the reaction plane, which
are qualitatively consistent with the CME. A clear signal compatible with a charge dependent
separation relative to the reaction plane is observed, which shows little or no collision energy
dependence when compared to measurements at RHIC energies. This provides a new insight for understanding
the nature of the charge-dependent azimuthal correlations observed at RHIC and LHC energies.

Recent years, lots of attentions~\cite{lab21,lab22,lab23,lab24,lab25} have been paid to the
chiral magnetic effect (CME).  It is shown that this effect originates from the existence of nontrivial topological
configurations of gauge fields and their interplay with the chiral anomaly which results in an asymmetry between
left- and right-handed quarks.  The created strong magnetic field
coupled to a chiral asymmetry can induce an electric charge current along the direction of a magnetic field. The strong magnetic field will
separate particles of opposite charges with respect to the
reaction plane. Recently, possible CME and topological charge
fluctuations have been recognized by
QCD lattice calculations in  gauge theory~\cite{lab26,lab27} and in QCD + QED with dynamical $2 + 1$ quark
flavors ~\cite{lab28}. Thus, such topological and CME effects in QCD might
be recognized in relativistic heavy-ion collisions directly in the presence
of very intense external electromagnetic fields.

Lots of analytical and numerical calculations indicate existence of extremely powerful electromagnetic fields in relativistic
heavy-ion collisions~\cite{lab1,lab4,lab6,lab29}. They are the strongest electromagnetic fields that exist in nature~\cite{lab1,lab4,lab6,lab29}.
Ref.~\cite{lab30,lab31} has discussed the electromagnetic response of the plasma produced by relativistic heavy-ion collisions.
It is found that the effects to have an important impact on the field dynamics. An exact analytical and numerical solution for the space and time
dependencies of an electromagnetic field produced in heavy-ion collisions was presented in Ref ~\cite{lab32}. It was confirmed that nuclear matter plays a crucial role~\cite{lab33} in
its time evolution.

In Ref.~\cite{lab34,lab35}, we used the Wood-Saxon nucleon distribution
instead of uniform distribution to improve the calculation
of the magnetic field of the central point for non-central collision  in the RHIC and LHC energy
regions. In this paper, we will use the improved magnetic field model to calculate the spatial distribution feature of
the chiral magnetic field in the RHIC and LHC energy regions. The dependencies of the spatial features of magnetic fields
on the collision energies, centralities, and collision time will be systematically investigated, respectively.

The paper is organized as follows. The key points of the improved model of magnetic field are described in Sec. II.
The calculation results of the  magnetic field are present in Sec. III. A summary is given in Sec. IV.

\section{The Improved model of chiral Magnetic field}
The improved model of  magnetic field mainly contains three parts:

\noindent (1) As shown in Fig.1, two similar relativistic heavy nuclei with charge $Z$ and radius $R$ are traveling
in the positive and negative $z$ direction with rapidity $Y_0$.
At $t=0$ they go through a non-central collision with impact parameter $b$ at the origin point.
The center of the two nuclei are taken at $x=\pm b/2$ at  time $t=0$ so that the direction of $b$ lies along the $x$ axis.
The region in which the two nuclei overlap contains the participants, the regions in which they do not overlap contain
the spectators.

\begin{figure}[h!]
\centering \resizebox{0.5\textwidth}{!}{
\includegraphics{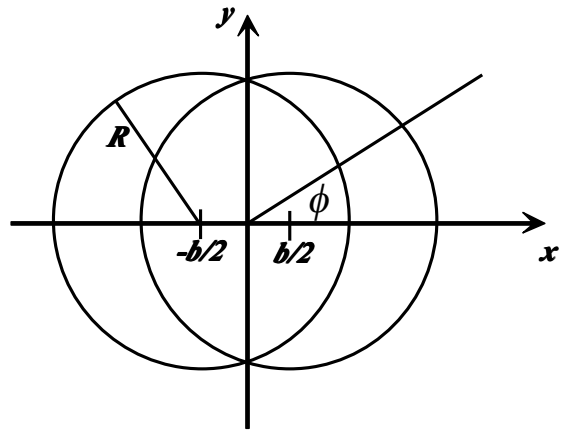}}
\caption{Cross-sectional view of a non-central relativistic heavy-ion collision along the $z$ axis.
            The two  nuclei have same radii $R$, move in opposite directions, and collide with impact parameter $b$.
            The angle $\phi$ is an azimuthal angle with respect to the reaction plane. The plane $y=0$ is called the reaction plane.
            The two nuclei overlap region contains the participants,and they do not overlap regions contain the spectators.}
\label{fig1}
\end{figure}

As the nuclei are nearly traveling with the speed of light in ultra-relativistic heavy-ion collision experiments,
the Lorentz contraction factor $\gamma$ is so large that the two included nuclei can be taken as pancake shape(as the $z = 0$ plane).
We use the Wood-Saxon nuclear distribution instead of uniform nuclear distribution~\cite{lab1}. The Wood-Saxon nuclear distribution forms is:
\begin{eqnarray}
n_A(r)=\frac{n_0}{1+\exp{(\frac{r-R}{d})}},
\label{eq:eq1} 
\end{eqnarray}

\noindent here $d$ = 0.54 fm, $n_0$ = 0.17 fm$^{-3}$ and the radius $R$=1.12 A$^{1/3}$ fm. Considering the Lorentz contraction,
the density of the two-dimensional plane can be given by:

\begin{eqnarray}
\rho_{\pm}(\vec{x}^\prime_\bot)=N\cdot\int_{-\infty}^{\infty}dz'\frac{n_0}{1+\exp(\frac{\sqrt{(x'\mp{b/2})^2+y'^{2}+z'^{2}}-{\rm R}}{d})},
\label{eq:eq2} 
\end{eqnarray}

\noindent where $N$ is the normalization constant. The number densities of the colliding nuclei can be normalized as

\begin{eqnarray}
\int{d}\vec{x}^\prime_\bot\rho_{\pm}(\vec{x}^\prime_\bot)=1.
\label{eq:eq3} 
\end{eqnarray}

\noindent (2) Secondly, in order to study the strength of the magnetic field caused by the two relativistic  traveling nuclei,
we can split the contribution of particles to the magnetic field in the time $t>0$. The specific forms of expression for the contribution of
particles to the magnetic field in the following way

\begin{eqnarray}
\vec{B}=\vec{B}^+_s+\vec{B}^-_s+\vec{B}^+_p+\vec{B}^-_p
\label{eq:eq4} 
\end{eqnarray}

\noindent where $\vec{B}^\pm_s$ and $\vec{B}^\pm_p$ are the the contributions of the spectators and the participants
moving in the positive or negative $z$ direction, respectively. For spectators, we assume that they do not scatter at all
and that they keep traveling with the beam rapidity $Y_0$. Combining with Eq.(2), we use the density above and give

\begin{eqnarray}
\lefteqn{e\vec{B}^\pm_s(\tau,\eta,\vec{x}_\bot)=\pm Z\alpha_{EM}\sinh(Y_0\mp\eta)
\int{d}^2\vec{x}^\prime_\bot\rho_{\pm}(\vec{x}^\prime_\bot)}\nonumber\\
&&\times[1-\theta_\mp(\vec{x}^\prime_\bot)]\frac{(\vec{x}^\prime_\bot-\vec{x}_\bot)\times\vec{e}_z}
{[(\vec{x}^\prime_\bot-\vec{x}_\bot)^2+\tau^2\sinh(Y_0\mp\eta)^2]^{3/2}},
\label{eq:eq5} 
\end{eqnarray}

\noindent where $\tau=(t^2-z^2)^{1/2}$ is the proper time, $\eta=\frac{1}{2}\ln[(t+z)/(t-z)]$ is the space-time rapidity, and

\begin{eqnarray}
\theta_\mp(\vec{x}^\prime_\bot)=\theta[R^2-(\vec{x}^\prime_\bot\pm\vec{b}/2)^2].
\label{eq:eq6} 
\end{eqnarray}

In the other hand, the distribution of participants that remain traveling along the beam axis is given by
\begin{eqnarray}
f(Y)=\frac{a}{2\sinh(aY_0)}{\rm e}^{aY},  \hskip1cm -Y_{0}\leq{Y}\leq{Y_{0}}.
\label{eq:eq7} 
\end{eqnarray}

\noindent Experimental data gives  $a\approx1/2$, which is consistent with the baryon junction stopping mechanism. The contribution
of the participants to the magnetic field can be  given by

\begin{eqnarray}
e\vec{B}^\pm_p(\tau,\eta,\vec{x}_\bot)=\pm Z\alpha_{EM}\int{\rm d}^2\vec{x}^\prime_\bot
\int{\rm d}Y f(Y)\sinh(Y\mp\eta)\nonumber\\
\times\rho_{\pm}(\vec{x}^\prime_\bot)\theta_\mp(\vec{x}^\prime_\bot)
\frac{(\vec{x}^\prime_\bot-\vec{x}_\bot)\times\vec{e}_z}
{[(\vec{x}_\bot^\prime-\vec{x}_\bot)^2+\tau^2\sinh(Y\mp\eta)^2]^{\frac{3}{2}}}
\label{eq:eq8} 
\end{eqnarray}

(3)  In the third part, in order to study the spatial distribution of the magnetic field, we will calculate the $eB_{x}$ and $eB_{y}$ components of the chiral magnetic field from
spectator and participant nuclei. The specific forms of the contribution of $eB_{x}$ and $eB_{y}$ components from the spectator and participant nuclei are given as follows:
\begin{eqnarray}
\lefteqn{eB^\pm_{sy}(\tau,\eta,\vec{x}_\bot)=\mp Z\alpha_{EM}\sinh(Y_0\mp\eta)
\int{d}^2\vec{x}^\prime_\bot\rho_{\pm}(\vec{x}^\prime_\bot)}\nonumber\\
&&\times[1-\theta_\mp(\vec{x}^\prime_\bot)]
\frac{(x^\prime-x)}
{[(\vec{x}^\prime_\bot-\vec{x}_\bot)^2+\tau^2\sinh(Y_0\mp\eta)^2]^{3/2}},
\label{eq:eq9} 
\end{eqnarray}

\noindent where $eB_{sy}$ is the $y$ component of magnetic field from spectators, and the $x$ component of magnetic field from spectators
is given by:
\begin{eqnarray}
\lefteqn{eB^\pm_{sx}(\tau,\eta,\vec{x}_\bot)=\pm Z\alpha_{EM}\sinh(Y_0\mp\eta)
\int{d}^2\vec{x}^\prime_\bot\rho_{\pm}(\vec{x}^\prime_\bot)}\nonumber\\
&&\times[1-\theta_\mp(\vec{x}^\prime_\bot)]
\frac{(y^\prime-y)}
{[(\vec{x}^\prime_\bot-\vec{x}_\bot)^2+\tau^2\sinh(Y_0\mp\eta)^2]^{3/2}},
\label{eq:eq10} 
\end{eqnarray}

In the other hand, the $y$ component of  magnetic field from participants is given by:
\begin{eqnarray}
eB^\pm_{py}(\tau,\eta,\vec{x}_\bot)=\mp Z\alpha_{EM}\int{\rm d}^2\vec{x}^\prime_\bot
\int{\rm d}Y f(Y)\sinh(Y\mp\eta)\nonumber\\
\times\rho_{\pm}(\vec{x}^\prime_\bot)\theta_\mp(\vec{x}^\prime_\bot)
\frac{(x^\prime-x)}
{[(\vec{x}_\bot^\prime-\vec{x}_\bot)^2+\tau^2\sinh(Y\mp\eta)^2]^{\frac{3}{2}}}
\label{eq:eq11} 
\end{eqnarray}

\noindent and the $x$ component of  magnetic field from participants is given by:
\begin{eqnarray}
eB^\pm_{px}(\tau,\eta,\vec{x}_\bot)=\pm Z\alpha_{EM}\int{\rm d}^2\vec{x}^\prime_\bot
\int{\rm d}Y f(Y)\sinh(Y\mp\eta)\nonumber\\
\times\rho_{\pm}(\vec{x}^\prime_\bot)\theta_\mp(\vec{x}^\prime_\bot)
\frac{(y^\prime-y)}
{[(\vec{x}_\bot^\prime-\vec{x}_\bot)^2+\tau^2\sinh(Y\mp\eta)^2]^{\frac{3}{2}}}
\label{eq:eq12} 
\end{eqnarray}

\section{The calculation results}

In order to study the dependencies of magnetic field $eB$ on proper time, we show the dependencies of magnetic field $eB$ (at central point $(x, y) = (0, 0)$)
on proper time $\tau$ at $\sqrt{s_{NN}}$ = 200 GeV for Au - Au collisions with b=8fm and  $\sqrt{s_{NN}}$ = 2760 GeV for Pb - Pb collisions with b = 8 fm, respectively. From Fig.2(a, b),
one can find that at small proper time the magnetic field is mainly from the contribution of spectator nucleons, but as the proper time increases, more and more large contribution of the magnetic field is from participant nucleon. Figure 2(c,d) show the comparisons of the magnetic field and the ratio of $(eB)_{p}/(eB)$ at $\sqrt{s_{NN}}$ = 200 GeV
and $\sqrt{s_{NN}}$ = 2760 GeV. One can find that at smaller proper time $\tau$($\tau<8\times 10^{-3}$ fm) the magnetic field at $\sqrt{s_{NN}}$ = 2760 GeV is greater than
that of $\sqrt{s_{NN}}$ = 200 GeV,
but when $\tau> 8 \times 10^{-3}$ fm, the magnetic field at $\sqrt{s_{NN}}$ = 2760 GeV is less than that of $\sqrt{s_{NN}}$ = 200 GeV. From Fig.2(d) one can find that the contribution of magnetic field from participant nucleons increases with the increase of proper time.

\begin{figure}[h!]
\centering \resizebox{0.48\textwidth}{!}{
\includegraphics{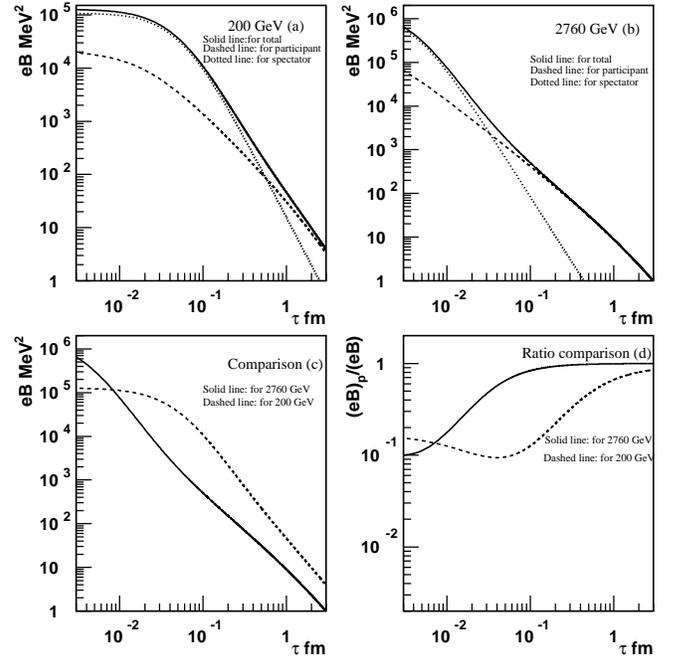}}
\caption{The dependencies of magnetic field $eB$ (at central point $(x, y) = (0, 0)$ )on proper time $\tau$ at $\sqrt{s_{NN}}$ = 200 GeV(a) for Au-Au collisions with b = 8 fm
and  $\sqrt{s_{NN}}$ = 2760 GeV(b) for Pb - Pb collisions with b = 8 fm, respectively. The dashed line is for the contribution from participant nucleons, and dotted
line is for the spectator nucleons.  The solid line is the combined contributions of participant and spectator nucleons.
(c) is for the comparison of the magnetic field between the two collision energies with proper time. (d) is for the comparison of ratio between the magnetic field from participant and the total magnetic field.}
\label{fig2} 
\end{figure}

Figure 3 shows the dependencies of magnetic field $eB$ (at central point $(x, y)=(0, 0)$ )on central of mass energy $\sqrt{s_{NN}}$ at different proper time
$\tau$. It is argued that at smaller proper time ($\tau$ = 0.001 and 0.0001fm) the magnetic fields increase with the increase of the CMS energy ($\sqrt{s_{NN}}$), but
with the increase of proper time ($\tau$), the magnetic field decreases sharply with increasing collision energy of  $\sqrt{s_{NN}}$. It is found that when
$\tau$ = 3 fm and $\sqrt{s_{NN}} > 200$ GeV, the magnetic field approaches zero.

\begin{figure}[h!]
\centering \resizebox{0.48\textwidth}{!}{
\includegraphics{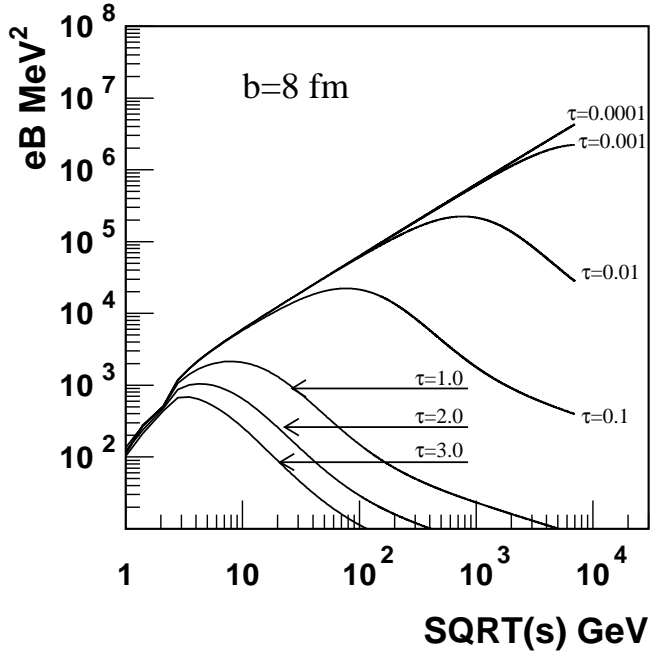}}
\caption{The dependencies of magnetic field $eB$ (at central point $ (x, y) = (0, 0)$ )on central of mass energy $\sqrt{s_{NN}}$ at different proper time
$\tau$ = 0.0001, 0.001, 0.01, 0.1, 1.0, 2.0 and 3.0 fm,respectively.}
\label{fig3} 
\end{figure}

For consistency with the experimental results, we take Au-Au collision with RHIC energy region and Pb-Pb collision with LHC energy region.
When studying the spatial distribution characteristics of magnetic field, we choose the spatial regions of
-10.0 fm $\leq x\leq$ 10.0 fm and -10.0 fm $\leq y\leq$ 10.0 fm.

\begin{figure}[h!]
\centering \resizebox{0.48\textwidth}{!}{
\includegraphics{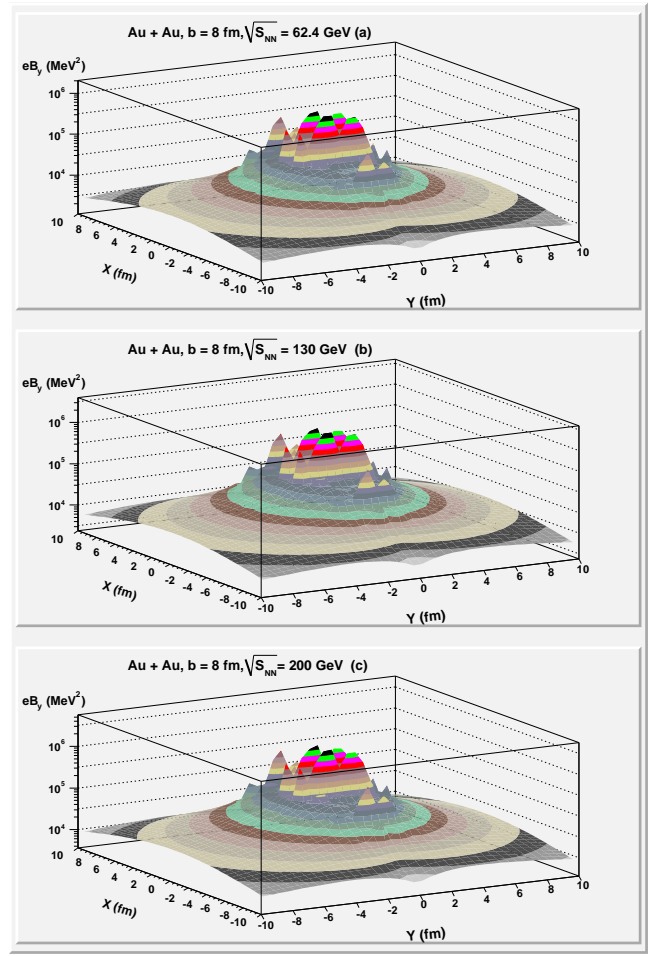}}
\caption{The dependencies of magnetic field spatial distributions of $eB_{y}$ on different collision energies $\sqrt{s_{NN}}$ = 62.4 GeV(a), 130 GeV(b) and 200 GeV(c), respectively.
The impact parameters $b = 8$ fm and proper times $\tau$ = 0.0001 fm.}
\label{fig4} 
\end{figure}


Figure 4 shows  the magnetic field spatial distributions of $eB_{y}$  with different collision energies $\sqrt{s_{NN}}$ = 62.4 GeV, 130 GeV and 200 GeV and proper time $\tau = 0.0001$ fm.
The collision energies shown in Fig. 4 are in RHIC energy region. The spatial distributions of $eB_{y}$ show obviously axis symmetry characteristics along $x = 0$ and $y = 0$ axes.
There is a peak around central point $(x, y) = (0, 0)$, and the magnetic field get smaller and smaller when the location go farther away from the center position.
When $\tau = 0.0001$ fm,

On both sides of $y = 0$ line, there are two symmetrical peaks.  These two peaks are almost connected when $\sqrt{s_{NN}}$ = 62.4 GeV. As the collision energy increases, the two peaks start to separate and expose the valley between the two peaks when $\sqrt{s_{NN}}$ = 130 GeV and 200 GeV. The maximum of magnetic field  $eB_{y}$ in RHIC energy region reaches $2.2\times 10^{5}$ MeV$^{2}$.

\begin{figure}[h!]
\centering \resizebox{0.48\textwidth}{!}{
\includegraphics{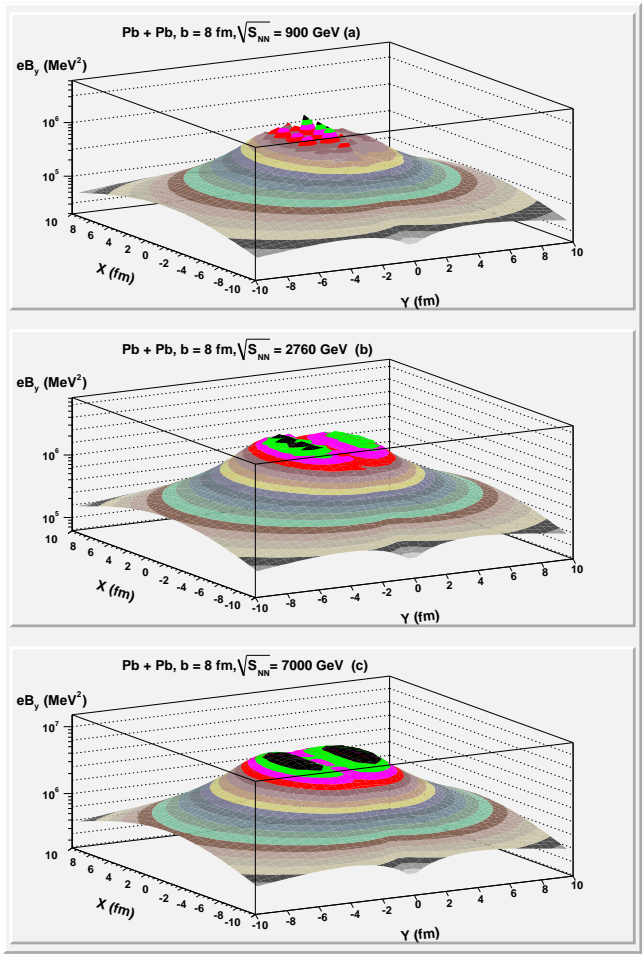}}
\caption{The dependencies of magnetic field spatial distributions of $eB_{y}$ on different collision energies $\sqrt{s_{NN}}$ = 900 GeV(a), 2760 GeV(b), 7000 GeV(c), respectively.
The impact parameters $b = 8$ fm and proper times $\tau$ = 0.0001 fm.}
\label{fig5} 
\end{figure}

Compared with Fig.4, Fig.5 shows the magnetic field spatial distributions of $eB_{y}$  in the LHC energy region. When the collision energy rises up to 900 GeV in LHC energy region, the distribution features of magnetic field have some differences from that of the RHIC energy region.  For example the magnetic field distribution peak around $x = 0$ and $y = 0$  becomes flat at $\sqrt{s_{NN}}$ = 900 GeV, and begin to appear the phenomenon of two peaks.  The maximum of magnetic field  $eB_{y}$ in LHC energy region reaches $2.0\times 10^{6}$ MeV$^{2}$, which is larger than that of RHIC energy region at $\tau$ = 0.0001 fm.

From Fig.2 to Fig.5, we argue that the magnetic field spatial distributions of $eB_{y}$ are highly inhomogeneous.
The distribution features in the RHIC energy region is different from that of the LHC energy region.
It is argued that at smaller proper time ($\tau$ = 0.001 and 0.0001fm) the magnetic fields increase with the increase of the CMS energy ($\sqrt{s_{NN}}$), but
with the increase of proper time ($\tau$), the magnetic field decreases sharply with increasing collision energy of central of mass $\sqrt{s}$.

The above we make a discussion of magnetic field spatial distributions with the collision energy and impact parameter relations, we will make a study of magnetic field with the
proper time.  The magnitude of magnetic field is presented as:

\begin{eqnarray}
eB=\sqrt{(eB_{x})^{2}+ (eB_{y}^{2})}
\label{eq:eq13} 
\end{eqnarray}

Sometimes, one often takes the $y$ component $eB_{y}$ to approximately replace $eB$.  This is the reason that $eB_{y}$ is usually larger than $eB_{x}$.
In order to verify the rationality of the substitution, we need a detailed study the relation between $eB_{y}$ and $eB$. Figure 6 shows the dependencies of the ratio of $eB_{y}/(eB)$ on  $x$ and  $y$  at  $\sqrt{s_{NN}}$=  200 GeV and at different proper time  $\tau$ = 0.02, 0.2 and 2.0 fm,
respectively. The Fig.6(a, c and e) are for $eB_{y}/(eB)$  with $y$ at different proper time. From Fig.6(a, c and e), one can figure out that the ratio of
$eB_{y}/(eB)$  with $y$ change is between 0.9 to 1.0. In this case, one can approximate the $eB_{y}$ instead of $eB$. Compared with the
relation of ratio $eB_{y}/(eB)$  with $y$, the relationship of ratio $eB_{y}/(eB)$  with $x$ shown as Fig.6(b, d and f) is obviously different.
The main different is the dip located at $x = 0$. The minimum value of the ratio at $x = 0$  can be decreased to 0.5.

\begin{figure}[h!]
\centering \resizebox{0.5\textwidth}{!}{
\includegraphics{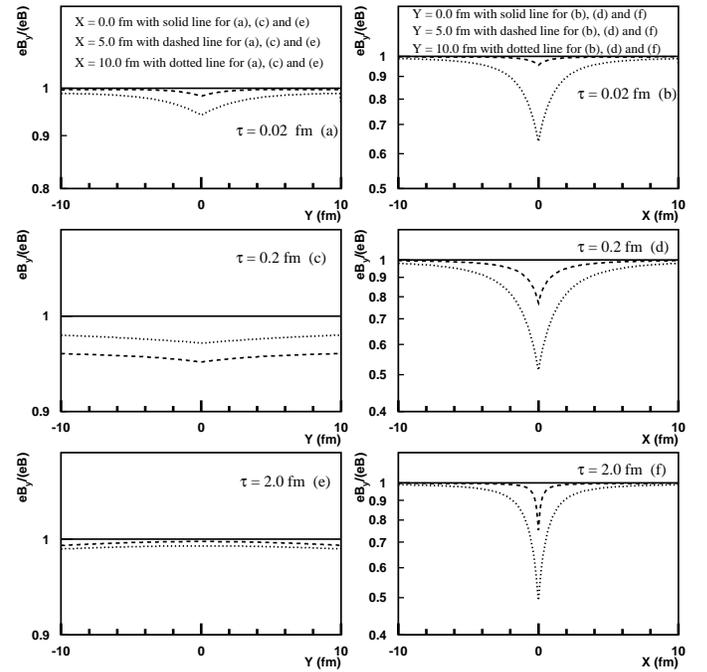}}
\caption{The dependencies of the ratio of $eB_{y}/(eB)$  on  $x$ and  $y$  at $\sqrt{s_{NN}}$=  200 GeV and at different proper time  $\tau$ = 0.02, 0.2 and 2.0 fm, respectively.
The Fig.6(a, c and e) are for $eB_{y}/(eB)$  with $y$ and Fig.6(b, d and f) are for $eB_{y}/(eB)$  with $x$ at different proper time.}
\label{fig6} 
\end{figure}

In order to study the spatial distribution of magnetic field on proper time, we show the dependencies of magnetic field $eB_{y}$ and $eB_{x}$ (at points $(x,y) = (5,5)$
and $(x, y) = (10, 10)$ ) on proper time $\tau$ at $\sqrt{s}$= 200 GeV for Au-Au collisions with b=8fm and  $\sqrt{s}$ = 2760 GeV and 7000 GeV for Pb - Pb collisions with b = 8fm, respectively. From Fig.7, one can find that at small proper time the the magnetic field increases with the increase of the collision energy, but the magnetic field of $\sqrt{s}$ = 7000 GeV decrease more quickly than that of $\sqrt{s}$ = 200 GeV with the increase of proper time. Fig.7(c,d) show that there is a relatively flat region with proper time  at point $(x, y) = (10,10)$ than that at point $(x,y) = (5,5)$.

\begin{figure}[h!]
\centering \resizebox{0.5\textwidth}{!}{
\includegraphics{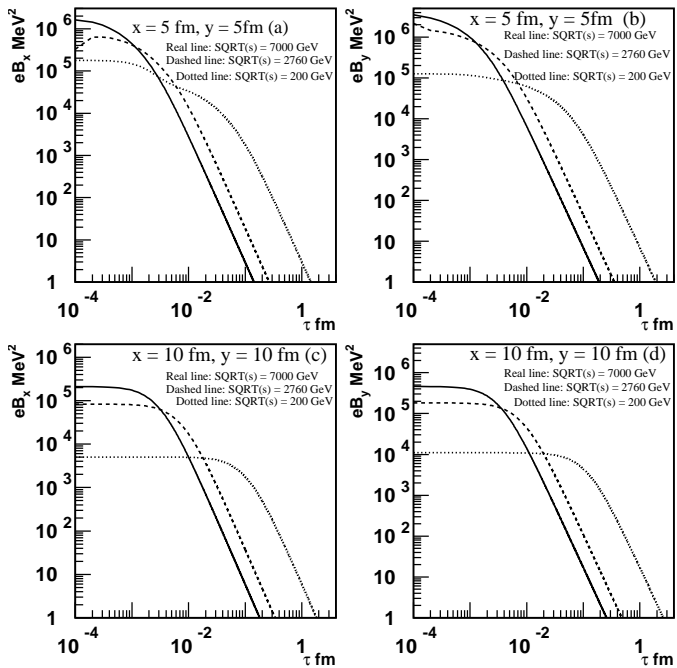}}
\caption{The dependencies of the magnetic field($eB_{x}$ and $eB_{y}$) on proper times $\tau$ at $(x, y) = (5, 5)$ and $(10, 10)$ at $\sqrt{s_{NN}}$ = 7000, 2760 and 200 GeV
respectively. The solid line is $\sqrt{s_{NN}}$ = 7000 GeV, the dashed line is for $\sqrt{s_{NN}}$ = 2760 GeV and dotted
line is for $\sqrt{s_{NN}}$ = 200 GeV.}
\label{fig7} 
\end{figure}

\section{Summary and Conclusion}

It is shown that an enormous magnetic field can indeed
be created in off-central heavy-ion collisions.  The magnetic field distributions of $eB_{x}$ and $eB_{y}$ are highly inhomogeneous, and $eB_{x}$ and
$eB_{y}$ distributions are completely different.  The enormous magnetic field is produced just after the collision, and the magnitude of magnetic field of LHC energy region
is larger that of RHIC energy region at the small proper time($\tau < 8.0 \times 10^{-3}$ fm). We are really surprised to find that the magnetic field in the LHC energy region decreases more quickly with the increase of the proper time than that of RHIC energy region. As the proper time $\tau$ increases to a certain value $8.0 \times 10^{-3}$ fm, the magnitude of magnetic field in the RHIC energy region begin to be larger than that of LHC energy region.

The dependencies of the ratio of $eB_{y}/(eB)$ on  $x$ and  $y$  at  different collision energies at RHIC and LHC and at different proper time are analyzed in this paper.
In most cases, the ratio $eB_{y}/(eB)$ approaches $1$, so this is a good approximate by using  $eB_{y}$ to approximately replace $eB$.  But one should note
that the ratio $eB_{y}/(eB)$ is between $0.5 \sim 1.0$ along $x = 0$ line.

We systematically study the spatial distribution features of chiral magnetic field in relativistic heavy-ion collisions at energies
reached at LHC and RHIC  with the improved model of chiral magnetic field in this paper.   The feature of chiral magnetic fields
at  $\sqrt{s_{NN}}$= 900, 2760 and 7000 GeV in the LHC energy region and  $\sqrt{s_{NN}}$ = 62.4, 130 and 200 GeV in the RHIC energy region
are systematically studied. 

The dependencies of the magnetic field on proper time for
at RHIC and LHC energy regions, respectively.  Comparing with that of RHIC energy region, one finds that the magnitudes of the magnetic fields with
proper time fall more rapidly at LHC energy region. The variation characteristics of magnetic field with impact parameter at RHIC energy region are
different from that of LHC energy region.  The maximum position is located in the small proper time ($\tau \sim 0.0001$ fm), more off-central collisions  and
$\sqrt{s_{NN}}\sim 7000$ GeV. The maximum of magnetic field in our calculation is about $eB \simeq 2 \times 10^{7} MeV^{2}$ when $\tau = 0.0001$
$b\simeq 8 fm$ and $\sqrt{s_{NN}}\sim 7000$ GeV.

\section{Acknowledgments}
This work was supported by the National Natural Science
Foundation of China (Grants Nos. 11375069, 11435054, 11075061, and 11221504), also by the Open innovation fund of the Ministry
of Education of China under Grant No. QLPL2014P01.

{}

\end{document}